\documentclass[prb,preprint]{revtex4-1} 
\usepackage{amsmath}  % needed for \tfrac, \bmatrix, etc.
\usepackage{amsfonts} % needed for bold Greek, Fraktur, and blackboard bold
\usepackage{graphicx} % needed for figures

\def\source#1#2#3#4{{\it #1}~{\bf #2}, #3 (#4)}
\def\be{\begin{equation}}
\def\ee{\end{equation}}
\def\bea{\begin{eqnarray}}
\def\eea{\end{eqnarray}}
\def\Eq#1{Eq. \ref{#1}}
\def\Eqs#1#2{Eqs. \ref{#1} and \ref{#2}}
\def\ie{{\it i.e.}}
\def\eg{{\it eg}.}
\def\oort{\frac{1}{\sqrt{2}}}

\def\CC{\cal C}
\def\ket#1{| #1 \rangle}
\def\bra#1{\langle #1 |}
\def\kpsi{\ket{\psi}}
\def\kPsi{\ket{\Psi}}
\def\kPsio{\ket{\Psi_o}}

\def\trace{\hbox{Tr}}

\def\braket#1#2{\langle #1 | #2 \rangle}
\begin{document}

% Be sure to use the \title, \author, \affiliation, and \abstract macros
% to format your title page.  Don't use lower-level macros to  manually
% adjust the fonts and centering.

\title{Bell-EPR Correlations within Local Quantum Theory}
% In a long title you can use \\ to force a line break at a certain location.

%When submitting the manuscript for review, do not include the author's name or institution
\author{Jay Lawrence}
	\email{jay.lawrence.@dartmouth.edu}
	 \affiliation{Department of Physics and Astronomy, Dartmouth College, Hanover, NH, USA} 
%	  \affiliation{The James Franck Institute. The University of Chicago, Chicago, IL, USA}  
%\altaffiliation[permanent address: ]{101 Main Street, Anytown, USA} % optional second address

% See the REVTeX documentation for more examples of author and affiliation lists.

\date{\today}

\begin{abstract}
We present a local unitary theory of a Bell-EPR measurement, starting
with the premeasurement filtering of the individual photon polarizations and
extending through the detection process involving four photodetectors, two at 
each receiving station.  The essential feature is that decoherence occurs locally 
and independently with each detector upon its absorption of a photon.  
Communication between observers after they read their local outcomes
confirms the known Bell-EPR correlations.  This theory is manifestly local, 
but there exist other formulations, and interpretations, that are non-local.
\end{abstract}

\maketitle % title page is now complete

\section{Introduction} 

It is said that Bell's theorem \cite{Bell} and experiments based on it \cite{CHSH, 
Clauser2, Aspect} show that quantum mechanics is non-local.\cite{google}  This 
statement is literally not true.  What Refs. 1 - 4 do show is that a 
{\it hidden variables} formulation 
of quantum theory must be nonlocal, as is the Bohm theory.\cite{Bohm} But 
quantum theory in its more familiar and useful formulation \cite{Weinberg} is not 
a hidden variables theory; it relinquishes the property of {\it realism} (a property 
conferred by hidden variables, in which a measurement reveals a pre-existing 
value), for the possibility of a local formulation.  As explained by Neilsen and 
Chuang,\cite{NCnote,NC} any formulation must violate either realism or locality 
(or both), and so it is said to violate {\it local realism}.\cite{NCnote} In this paper 
we show that the most basic formulation of quantum theory which accounts 
explicitly for decoherence is a local theory, and we show that it reproduces the 
observed correlations.  However, there {\it are} other formulations, and 
interpretations, that render a nonlocal form.  Here are some illlustrative 
examples, but first - what do we mean by local?

\subsection{Local Quantum Theory}
%(somewhere we need to discuss the meaning of 
%''local'' and ``non-local,'' as applied to quantum phenomena)

We consider a theory local if (1) no nonlocal interaction occurs in its analytical
dynamics (\eg, Schr\"{o}dinger evolution in Hilbert space), and (2) no non-local 
influence occurs through a measurement (or collapse) postulate.  The 
following interpretations \cite{Weinberg2} illustrate both of these conditions.

\subsection{Dependence on interpretations}

In the standard textbook interpretation of quantum mechanics,
\cite{Weinberg,Sakurai} a non-local influence arises when the collapse 
postulate\cite{survival} is applied to the Bell-EPR scenario.  A common 
rationale\cite{google} goes as follows:  Suppose that Alice measures her qubit 
and finds the definite result, say, either 0 or 1.  She then knows immediately 
that Bob's measurement will yield the opposite result, and it is often said
that Alice's measurement collapsed the state of both her qubit and 
(simultaneously) Bob's qubit.  This typical argument for non-locality is 
flawed because the two local measurement events can have space-like 
separation, and hence no causal relationship.  In fact there is no causal 
relationship even with time-like separation, where the same 
correlations occur.

%This is clearly a non-local effect if one regards the state vector as
%applying to the system under study;

The situation is less straightforward in an observer-centric approach (like
the Copenhagen Interpretation), where one regards the state vector as
representing the information available to an observer, based on the available
resources. The collapse of the state vector then refers to an updating of the
observer's information.  So Alice's measurement result, together with her
knowledge of the prepared Bell state, provide information about both her
result and Bob's (anticipated) result - simultaneously.  But both pieces of 
information reside with Alice, and not with Bob.  Alice did not actually 
measure Bob's qubit - she is only making a prediction of what Bob will 
find.  The state of Bob's qubit is not actually {\it determined} until a local 
meaurement is performed - this follows Wheeler's dictum,\cite{Wheeler} 
``No phenomenon is a real phenomenon until it is an observed 
phenomenon.''  
%If this example muddies the waters, our analysis in Sec. III 
%will bring clarity.

The Objective Collapse Theories (OCT) \cite{OCT,GRW,Pearle,Bassi} make
quantum theory non-local by introducing model interactions which cause the 
collapse.  These interactions must be both nonlinear and stochastic, even in 
a one-particle measurement.   In the Bell-EPR scenario, they must collapse 
the state of two separated detectors, whose settings can be chosen after the 
photons have been emitted. It is difficult to see how such a model could result 
in the observed correlations without non-local influence on the detectors.  
%It would have to produce random but perfectly anticorrelated outputs 
%at spacelike separated points.  It 

The Everett relative state theory\cite{Everett} is based on unitary evolution 
under known (local) interactions, and the local branching of an observer's 
memory state incorporating all possible outcomes. On each branch, the 
observer is unaware of all but the single outcome displayed on that branch. 
This theory may be applied to the Bell-EPR scenario, with two observers, 
and it reproduces the observed correlations without introducing non-locality.  
It should be noted that Everett himself did not make the Many-Worlds 
Interpretation; this came later from DeWitt and Graham.\cite{Graham} 

%the Everett conclusions are consistent with those of decoherence theory, 
%which will be described below.  
%supporting Everett's insight that an observer is blind to alternate outcomes
%CAI**[[TOO SPECIFIC HERE Further developments aligned with the 
%Everett theory by Mordecai Waegell \cite{Cai} involve the local branching 
%at measurements by individual observers.  This theory respects relativistic 
%causality and is manifestly local.

Each of the interpretations/formulations above is consistent with what 
we know from experiment.  This ambiguity reflects our ignorance 
regarding underlying foundational questions.  For example, 
we do not know whether quantum dynamics is unitary at the deepest 
level. Since the question of locality depends on interpretation, we do 
not know whether quantum mechanics is intrinsically local or non-local.
Our goal in this work is simply to describe a local {\it formulation} of
quantum theory, and to show that this is capable of reproducing the
correlations which are sometimes attributed to non-locality.  We believe 
that the {\it possibility} of a local formulation is important both for 
understanding the foundations of quantum theory, and in searching 
for unification with gravity.  

\subsection{Other Opinions}

In contradistinction to our position, most authors argue for locality {\it or} 
non-locality, and not for a choice based on formulation.  Both positions are well 
represented in the literature.  For example, a compelling argument in favor of
locality is given by R. Griffiths through a proof\cite{Grif1} and later extension 
of arguments.\cite{Grif2}  His approach utilizes the consistent histories 
formulation of quantum theory. In the other direction, there is an excellent 
review\cite{Brunner} which asserts that quantum theory is non-local, and it
discusses a broad range of phenomena in these terms.  However, in the 
presentation of the theory, it is noted that the term locality is interpreted more 
broadly to include local realism.\cite{NCnote}  As this example illustrates, there 
is no universal adherence to a common understanding of a precise meaning for 
the term non-locality; it is often used to characterize physical effects, and 
(less often) quantum theory itself, so that one should be prepared to 
interpret the term in the context of the discussion at hand.

Here is an outline for the rest of the paper:  In the next section we trace the 
dynamics from the photons to the detectors.  In Sec. III we account for 
decoherence due to the detectors' internal degrees of freedom and to the 
external environment, and we describe the resulting outcome as seen by 
observers.  The calculation raises foundational issues which we discuss in
Sec. IV, and we then conclude in Sec. V.
 
\section{The Evolution of States}

We imagine a pair of entangled photons moving away from each other (one to 
the left and the other to the right, as denoted by subscripts) in the singlet 
polarization state, 
\be
     \kpsi = \oort \big(\ket{V}_L \ket{H}_R - \ket{H}_L \ket{V}_R \big),
\label{photons1}
\ee
where $V$ and $H$ refer to individual vertical and horizontal polarizations.  Each 
photon encounters a  polarizing beam splitter, as shown in Fig. 1, oriented to 
separate $V$ and $H$ components and direct each to the appropriate detector.  
Each detector has two pointer states,\cite{pointer} called 1 if it receives the 
photon, and 0 if not.  So the initial state of the four-detector system (its ``ready 
state'') is 
\begin{figure}[h!]
\includegraphics[scale=0.65]{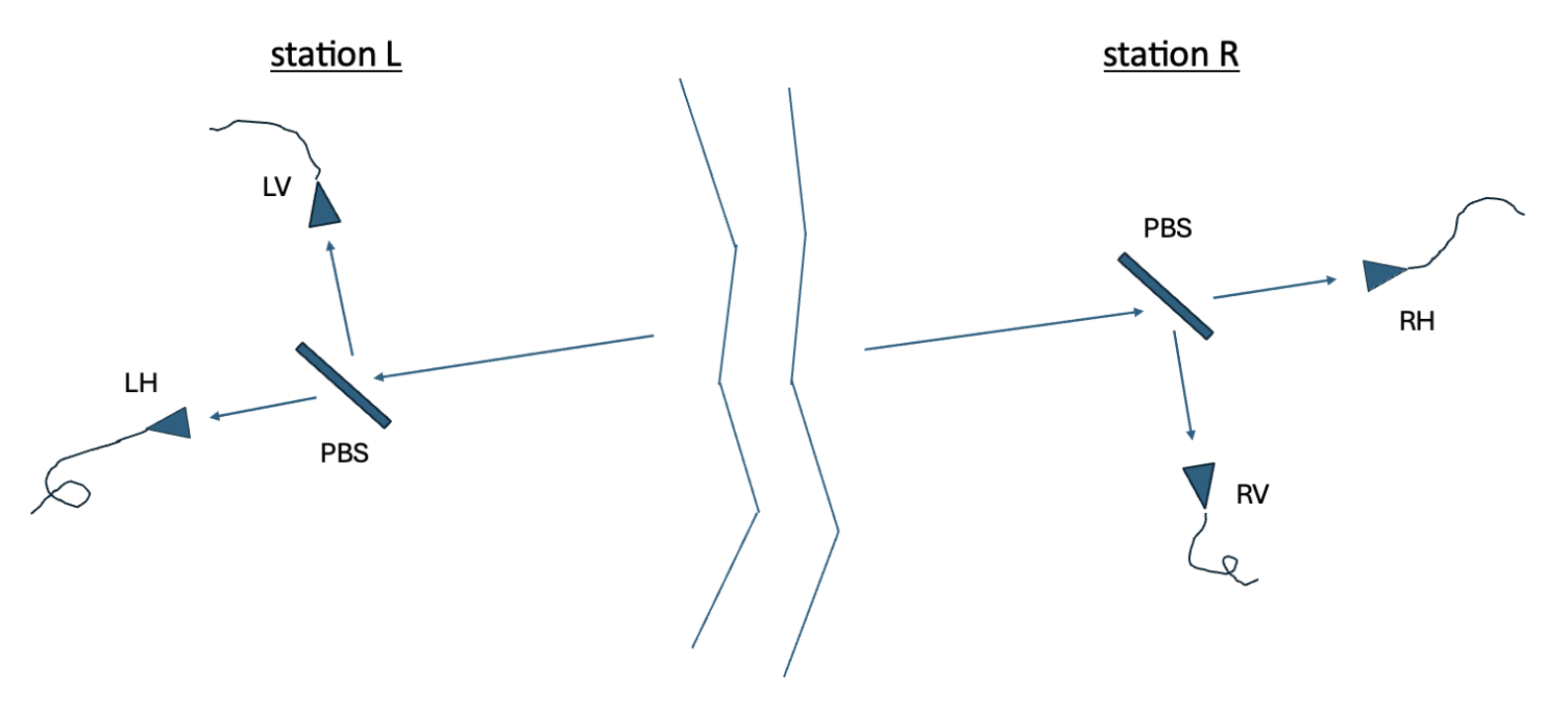}
\caption{\label{fig1} Far-separated detector systems for measuring the 
polarization of each photon.  At each station there are detectors for both 
vertical and horizontal polarizations.}
\end{figure}
\be
   \kPsio = \ket{0}_{LV} \ket{0}_{LH} \ket{0}_{RV} \ket{0}_{RH},
\label{detectors1}
\ee
where subscripts identify the detector by its position - its global position  
($L$ or $R$) and its more local position  ($V$ or $H$), for the specific 
polarization pathway along which it lies. The initial state of the joint system of 
photons and detectors is the tensor product 
of \Eqs{photons1}{detectors1}.  In the first term, the left-moving photon will be 
absorbed by the $LV$ detector while the right-moving photon will be absorbed 
by $RH$.  In the second term, the left-moving photon will be absorbed by the 
$LH$ detector and the right-moving by $RV$.  Upon the absorption of both
photons, the four-detector state therefore becomes
\be
   \kPsi = \oort \big( \ket{1}_{LV} \ket{0}_{LH} \ket{0}_{RV} \ket{1}_{RH} -
                \ket{0}_{LV} \ket{1}_{LH} \ket{1}_{RV} \ket{0}_{RH} \big).
\label{detectors2}
\ee
This has the form of a generalized Bell state, which mirrors the Bell state of the 
two photons.  But it is not so simple, because each detector has internal degrees 
of freedom, some of which are involved in the detection process, so that their states
change when a photon is absorbed.

So, reserving the labels ($0,1$) for the pointer states, we add labels for the states 
of the internal degrees of freedom, as follows:  $\ket{0,\mu}$ labels a ``no-photon'' 
state, and $\ket{1,\mu'}$ a ``photon-absorbed'' state.  If the detector is a 
photomultiplier tube, then the relevant degrees of freedom are electrons:  $\mu$ 
involves bound electrons in the dynodes, and $\mu'$ involves freed elecrons, which
produce a current.  The (final) states $\mu'$ are unitarily evolved from $\mu$ states 
by many-body interactions within the detector if and when it absorbs a photon. Now, 
each detector has its own distinct sets of such internal states, so that the initial and 
final states of the detector {\it array} (\Eqs{detectors1}{detectors2}) become, 
respectively,
\be
    \kPsio = \ket{0,\mu}_{LV} \ket{0,\nu}_{LH} \ket{0,\sigma}_{RV} \ket{0,\tau}_{RH},
    \hskip2truecm   \hbox{and}
\label{detectors3}
\ee
\be
    \kPsi = \oort \big( \ket{1, \mu'}_{LV} \ket{0, \nu}_{LH} \ket{0, \sigma}_{RV} 
    \ket{1, \tau'}_{RH} -  \ket{0, \mu}_{LV} \ket{1, \nu'}_{LH} \ket{1, \sigma'}_{RV} 
    \ket{0, \tau}_{RH} \big),
\label{detectors4}
\ee
where ($\mu,\nu,\sigma,\tau$) label the initial internal states associated with 
no-photon pointer states ($0$), while their primed counterparts 
($\mu',\nu',\sigma',\tau'$) label the corresponding final internal states 
associated with pointer states ($1$).   The detection process in the detector 
($LV$), for example, involves the transition from the metastable states ($0,\mu$) 
to stable states ($1,\mu'$) of that detector.  And although the transition is driven  
by unitary interactions, we do not have the control to reverse the evolution, 
so that the transition is irreversible thermodynamically.\cite{JL1}

The transition which is actually observed in the {\it detector array} takes us from 
states like \Eq{detectors3} to states like {\it one of the terms} in \Eq{detectors4},
- that is, from the pointer state ($0000$) to either ($1001$) or ($0110$).  The 
inability of the apparatus to display superpositions of the latter (macroscopically 
distinguishable) states is called the ``branching'' of the wavefunction - but this 
is more correctly described in terms of a density matrix,\cite{Taylor} as follows.

\section{The Density Matrix and Outcomes}

Because the detectors are macroscopic, the spectrum of internal states is 
dense, and pure states such as \ref{detectors3} and \ref{detectors4} are 
not realizable.  Although indeed we see definite pointer states, these are 
inevitably accompanied by mixtures of internal states, and the combination 
is appropriately described by density matrices.\cite{Zeh.70}  The density 
matrix of the four-detector array corresponds to an ensemble of which 
\Eq{detectors3} (or \ref{detectors4}) is a single element.  Thus the density 
matrix is the outer product, summed over states of the internal degrees 
of freedom, 
\be
     \rho = \frac{1}{2}\sum_{\mu \nu \sigma \tau} p_{LV}(\mu) p_{LH}(\nu) 
               p_{RV}(\sigma) p_{RH}(\tau) \ket{\Psi} \bra{\Psi},
\label{rho1}
\ee
where $\kPsi$ is given by \Eq{detectors3} in the initial state, and
\Eq{detectors4} in the final state; and $p_{LV}(\mu)$, etc., are the normalized 
probability distributions over the internal states of the individual detectors 
($LV$), etc. The product of individual probabilities expresses the independence 
of the detectors.  In the initial state, each probability factor could represent an 
approximately canonical distribution \cite{canonical,Popescu} with respect to 
the metastable equilibrium (the ``ready state'' $\ket{0}$) of the corresponding 
detector, resulting from interactions with its local external environment. 
The precise function is unimportant for our arguments. 

As the detector array transitions to its final state following the absorption of 
both photons, the internal states of two detectors evolve (for example, $\mu 
\rightarrow \mu'$, and $\tau \rightarrow \tau'$ in the first branch of 
\ref{detectors4}) under the many-body interactions within the detectors ($LV$ 
and $RH$) and, to a lesser extent, under the weaker interactions with their 
external environments.  For simplicity, we assume that these state changes 
comprise the important dynamics within each detector, and we ignore the 
slower changes in the probability functions (\ie, re-equilibration), so that the 
same functions $p_{LV}$, etc., apply to the final states.  Correcting this 
assumption would not affect our results.  We may consider the final 
state of the array to be reached, effectively, when two detectors register the 
arrival of the photons, even though the slower process of re-equilibration 
may not be complete by then.  This is justified because the signaling of a 
detector is driven by the rapid evolution of the internal response, and this 
is clearly irreversible once the detector signals.

To extract what observers will see, we take the partial trace of $\rho$ 
(\Eq{rho1}) over the states of the internal degrees of freedom; this defines 
the reduced density matrix 
\be
    \rho^{(r)} \equiv \trace_{\mu\nu\sigma\tau} \rho,
\label{reddef}
\ee
in the Hilbert space of the pointer states.  Since there are two macroscopically 
distinct pointer states of the array
\big[write \Eq{detectors4} as $\kPsi = (\ket{A} + \ket{B})/\sqrt{2} \big]$, 
$\rho^{(r)}$ will be a $2 \otimes 2$ matrix with entries labeled by these pointer
states.  We write out the resulting matrix elements of $\rho^{(r)}$ below.  The 
diagonal elements are the {\it direct} outer products; the first is
\bea
    \rho_{AA}^{(r)} =   \frac{1}{2}  \bigg( \ket{1}_{LV} \sum_{\mu} p_{LV}(\mu) 
    \braket{\mu'}{\mu'}_{LV} \bra{1} \bigg) \otimes \bigg( \ket{0}_{LH} \sum_{\nu} 
    p_{LH}(\nu) \braket{\nu}{\nu}_{LH} \bra{0} \bigg) \otimes     \nonumber   \\
    \bigg( \ket{0}_{RV} \sum_{\sigma} p_{RV}(\sigma) \braket{\sigma}{\sigma}_{RV} 
    \bra{0} \bigg) \otimes  \bigg( \ket{1}_{RH} \sum_{\tau} p_{RH}(\tau) 
    \braket{\tau'}{\tau'}_{RH} \bra{1} \bigg),
 \label{red1}
 \eea 
which simplifies under the normalization of the probability distributions, to 
 \be
    \rho_{AA}^{(r)}  =  \frac{1}{2} \ket{1}_{LV} \bra{1} \otimes \ket{0}_{LH} \bra{0} \otimes
                             \ket{0}_{RV} \bra{0} \otimes  \ket{1}_{RH} \bra{1}.
\label{red1prime}
\ee
A similar calculation for the other diagonal element gives us
\be
   \rho_{BB}^{(r)} =  \frac{1}{2} \ket{0}_{LV} \bra{0} \otimes \ket{1}_{LH} \bra{1} \otimes
                           \ket{1}_{RV} \bra{1} \otimes  \ket{0}_{RH} \bra{0}.
   \label{red2}
\ee
 The off-diagonal elements are the {\it indirect} outer products,
 \bea
    \rho_{AB}^{(r)} = \frac{1}{2} \bigg( \ket{1}_{LV} \sum_{\mu} p_{LV}(\mu) 
    \braket{\mu}{\mu'}_{LV} \bra{0} \bigg)\otimes \bigg( \ket{0}_{LH} \sum_{\nu} 
    p_{LH}(\nu) \braket{\nu'}{\nu}_{LH} \bra{1} \bigg) \otimes     \nonumber   \\
    \bigg( \ket{0}_{RV} \sum_{\sigma} p_{RV}(\sigma) \braket{\sigma'}{\sigma}_{RV} 
    \bra{1} \bigg) \otimes \bigg( \ket{1}_{RH} \sum_{\tau} p_{RH}(\tau) 
    \braket{\tau}{\tau'}_{RH} \bra{0} \bigg),
\label{red3}      
\eea
and its Hermitian conjugate,  
\be
   \rho_{BA}^{(r)} = \rho_{AB}^{(r)*}.
\label{red4}
\ee
Equations. \ref{red1prime} - \ref{red4} describe what the pointers can display, and 
thereby what the observers are able to infer about the state (\ref{rho1}) itself.  The 
main point is that the off-diagonal elements are unobservably small (see below), as 
is typical of macroscopic systems.  This confirms that the array pointer may display 
either ($1001$) or ($0110$),  but {\it not} a superposition!   The first display
represents the two-photon state $\ket{V}_L \ket{H}_R$, the second  represents 
$\ket{H}_L \ket{V}_R$.  And indeed, one sees a random sequence of definite but
opposite outcomes at the two sites.  The definiteness of pointer readings marks 
the transition from quantum to effectively classical behavior in the measurement 
chain:  Further links, such as the reading of the pointer by a human observer or 
a robot, remain effectively classical in revealing only the same definite state as
displayed by the pointer.  

To demonstrate the smallness of the off-diagonal components in $\rho^{(r)}$, 
compare \Eq{red3} with the analogous expression (\ref{red1}) for a diagonal element.  
Whereas in (\ref{red1}) the inner products are all unity, in (\ref{red3}) the magnitudes 
$|\braket{\mu}{\mu'}|$ are typically much less than unity, the phases are random, and 
the summations cover enormous numbers of terms.  It should further be noted that
each {\it detector-specific} summation in (\ref{red3}) is by itself unobservably small.  
Each such summation may be interpreted as an off-diagonal element of the reduced 
density matrix of a {\it single detector}, where it prevents the appearance of 
superpositions of its {\it own} pointer states, 0 and 1.  This is the origin of effectively 
classical behavior in any properly functioning detector.

\subsection{Imperfect Correlations}

Throughout the paper we have assumed that Alice and Bob use the same
measurement settings.  The proof of Bell's theorem\cite{Bell} requires the 
use of different settings, which introduces imperfect correlations.  This
complication in no way compromises the locality of the theory.  For
completeness, we derive Bell correlations and confirm this point in
the Appendix. 

\section{Discussion}

%The following commentaries will help flesh out the meaning of the above results.

\subsection{Behind the Partial Trace}

%\be
%   \rho(t=0) = \frac{1}{2}\sum_{\mu \nu \sigma \tau} p_{LV}(\mu) p_{LH}(\nu) 
%               p_{RV}(\sigma) p_{RH}(\tau) \ket{\Psi(t=0)} \bra{\Psi(t=0)}
%\label{rho0}
%\ee
The {\it array} pointer states emerging from the partial trace, (1001) and (0110), 
comprise what is called the ``preferred basis'' of orthodox decoherence 
theory.\cite{conventional,Zurek.81,Schloss.04}  This term refers to the effectively
classical states which are selected by interactions between the pointer and the 
environment (the so-called ``environmentally-induced superselection''). 

The selection process originates in our case with the pointer states of an
individual detector (0 and 1).  These are selected by the (local) interactions 
with the internal degrees of freedom of {\it that} detector.  They correspond to 
different local minima of the detector's free energy, which give rise to different 
effective Hamiltonians for the corresponding internal degrees of freedom. 
%Superpositions of these pointer states are unstable under decoherence.
Given these detector-specific pointer states, the {\it array} pointer states are 
generated by the quantum dynamics of \Eq{detectors1} $\rightarrow$ 
\ref{detectors2}, starting from the initial state (0000).

The internal degrees of freedom comprise the most important part of the 
environment, but there is also the external part.  The relevant parts of this
are the {\it local} external environments - the parts which interact directly with 
a particiular detector.  In the tracing procedure, when we sum explicitly over 
the states of the internal degrees of freedom (\Eqs{rho1}{reddef}), we also 
account implicitly for the external interactions through the ensembles
 [$p_{LV}(\mu),...,p_{RH}(\tau)$] of \Eq{rho1}, which are determined by 
these interactions.  This distinction between internal and external 
environments, with their different roles in the measurement process, 
is the unique aspect of our treatment, allowing us to maintain locality 
in the account for decoherence.  

We emphasize that the partial trace does not introduce non-locality into 
the theory, despite the possible illusion that the correlated pointer readings 
at $L$ and $R$ emerged from it.  In fact the partial trace does not represent 
a physical process, but rather a limitation on information available to observers.  
The correlated pointer readings are already implicit in the microscopic states 
of (\ref{detectors4}) if one recognizes that, owing to the internal degrees of 
freedom, interference between the two pointer states cannot occur.  The 
partial trace makes this explicit in the reduced density matrix.

%The apparent collapse of the state vector as represented by 
%Eqs. \ref{red1prime} - \ref{red4} applies to this limited information, and not 
%to the physical state of \Eqs{detectors3}{rho1}.*******
 
 We address this issue more graphically by demonstrating Bell-EPR 
correlations from the full (unreduced) mixed state (\Eq{rho1}),
eliminating the need for the partial trace.  This should remove 
any confusion over the lack of simultaniety of detector signals. 

\subsection{Observers and Locality}

 Assume that Alice (at station L) and Bob (at station R) record what their 
detectors tell them - the polarization ($V$ or $H$) of the photon absorbed.  
They do this for each of $N$ measurements on identically prepared photon 
pairs, and then they compare lists.  They could do this by telephone or by 
meeting in person, but we automate the comparison as follows:  Place a 
robot at each station to read the output of, say, its $V$-detector.  Then 
encode this output in a variable $Z$, which takes the value ($+1$ or $-1$), 
for the output (0 or 1) respectively.  The $Z$-values from both stations are 
brought together and multiplied, and the resulting number 
($Z_{LV}Z_{RV}$) is read out and recorded by a third observer, Clara.  

It would now be natural to evaluate the expectation value, 
$\langle Z_{LV}Z_{RV} \rangle = \trace (\rho Z_{LV}Z_{RV})$, where 
the trace covers internal states as well as pointer states, but this only 
reproduces the reduced density  matrix in the process.  It is more 
instructive to simply observe that every element (\Eq{detectors4})
of the full ensemble (\Eq{rho1}) is an eigenstate of the operator
$Z_{LV}Z_{RV}$, with eigenvalue $(-1)$, and therefore
\be
     Z_{LV}Z_{RV} \rho = - \rho  \hskip1truecm  {\implies} \hskip1truecm
     \langle Z_{LV} Z_{RV} \rangle = -1.
\label{identity}
\ee
The perfect (anti)correlations show that both Alice's and Bob's entries 
($Z_{LV}$ and $Z_{RV}$) come from the same branch of the state vector 
at each measurement, so there is no ``interbranch'' communication.  This 
restriction is enforced by the  {\it local} inner products 
$\braket{\mu}{\mu'}$.  Of course, ``which branch'' the two entries come 
from varies randomly from one measurement to the next (as seen by
Alice and Bob), while from an ``outside perspective,'' the sequence of 
$N$ measurements generates $2^N$ branches of the state vector, 
representing all possible histories of the communication.  

\subsection{Against Interpretation}

This ``outside perspective'' corresponds to the Many-World's Interpretation, 
which we have not assumed, but which presents itself, inevitably, with unitary 
evolution.  The question of the actual survival of unobserved branches is
currently beyond the realm of experimental physics, and a matter up for
interpretations.  Strictly speaking, there is nothing to prevent one from 
reinserting the collapse axiom of the standard interpretation at this stage, 
in effect removing one of the branches at each measurement.  Then, only 
one of those $2^N$ possible histories would remain in the state.  However, 
in this interpretation the theory is no longer unitary, deterministic, nor local. 
Experimental physics has not yet supplied answers.

 \section{Conclusions}

We argued at the beginning that the existence of multiple viable interpretations of 
quantum theory, considered as reflections of our ignorance, allow for the possibility 
that quantum mechanics could be intrinsically either local or non-local.  

The paper is then devoted to demonstrating in detail how a local theory describes the 
observed Bell-EPR correlations.  The theory applies standard quantum mechanics 
assuming only known unitary interactions and based on no particular interpretation.
As in the most basic decoherence calculations,\cite{Zeh.70,Zurek.81,Schloss.04} 
it does not invoke the collapse axiom, as its goal is to account for 
measurement outcomes without it. 

In the Bell-EPR context, it is important to recognize that the interactions causing 
decoherence are local.  These include the interactions of the pointer with the 
internal degrees of freedom of a detector, and the interactions of these with the 
local external environments.  The latter enter the theory through the mixed-state 
ensembles defined in \Eq{rho1}, which are established through local interactions.
%We maintain locality, by accounting for the local external environment through 
%the mixed-state ensembles of \Eq{rho1}.

Owing to the unitarity of the theory, both of the possible outcomes are present
in the final state of the detector system, although observers can be aware of 
only one.  This effective classicality is traced to the individual detectors which, 
being macroscopic, can only display one of their ``allowed'' states, 0 or 1.  
Quantum dynamics then restricts the detector {\it array} to one of the 
states, (1001) or (0110). 

\acknowledgements

I would like to thank Mordecai Waegell for stimulating and enlightening
discussions.

\medskip

\noindent DECLARATION: The author has no conflicts to disclose.

\bigskip
\centerline{{\bf APPENDIX:  Imperfect Correlations}}

\bigskip
We generalize the analysis of Sec. III to account for different measurement settings by 
Alice and Bob. Let Alice (at $L$) use the ($V,H$)  basis while Bob (at $R$) uses a 
different basis (say $V',H'$), related to Alice's basis by a rotation through an arbitrary
angle $\theta$:
\bea
   \ket{V'} = \cos \theta \ket{V} + \sin \theta \ket{H},  \nonumber   \\
   \ket{H'} = - \sin \theta \ket{V} + \cos \theta \ket{H}.
\label{primed polarizations}
\eea
In these bases, the initial two-photon state (\ref{photons1}) has a\ four-term expansion
\be
  \ket{\psi} = \oort \bigg[ \cos \theta \big( \ket{V}_L \ket{H'}_R - \ket{H}_L \ket{V'}_R \big)
  + \sin \theta \big( \ket{V}_L \ket{V'}_R + \ket{H}_L \ket{H'}_R \big) \bigg],
\label{photons2}
\ee
and, after the absorption of both photons, the state of the detectors will also have a 
four-term expansion; the generalization of \Eq{detectors2} is
%(with internal degrees of freedom implicit) 
\bea
    \kPsi = \oort \bigg[ \cos \theta \big( \ket{1}_{LV} \ket{0}_{LH} \ket{0}_{RV'} \ket{1}_{RH'} -
                \ket{0}_{LV} \ket{1}_{LH} \ket{1}_{RV'} \ket{0}_{RH'} \big)  
                \nonumber  \\
               + \sin \theta \big( \ket{1}_{LV} \ket{0}_{LH} \ket{1}_{RV'} \ket{0}_{RH'} +
                \ket{0}_{LV} \ket{1}_{LH} \ket{0}_{RV'} \ket{1}_{RH'} \big) \bigg].
\label{detectors5}
\eea
The correlation function $\CC(\theta)$ is the probability (sin$^2 \theta$) of similar 
outcomes minus the probability (cos$^2 \theta$) of opposite outcomes, so that
 $\CC(\theta) = - \hbox{cos2} \theta$, which corresponds to the result found 
 experimentally.\cite{Aspect}

As above, the two photons communicate their entanglement to the detector system.  
But, also as above, the superposition of detector states involves the internal degrees 
of freedom, whose states depend on the displayed pointer states (0,1) and make the 
superposition incoherent.  So the state is again represented by a density matrix, which 
can be reduced by a partial trace over the internal degrees of freedom.  This leaves a 
$4 \times 4$ matrix whose off-diagonal elements are immeasurably small, and whose 
diagonal elements are constructed from \Eq{rho1}. This matrix is the $4 \times 4$ 
analog of the $2 \times 2$ defined by Eqs. \ref{red1} - \ref{red4}.  The matrix elements 
again depend only on the local inner products, $\braket{\mu}{\mu'}$, etc., so that no 
nonlocal influences are introduced by the dynamics.


\begin{thebibliography}{99}
\bibitem{survival} The collapse postulate is the stipulation that a measurement throws the    
       system into an eigenstate of the measured observable (Sakurai\cite{Sakurai} p.23).
\bibitem{pointer} The physical pointer could refer to a mechanical object taking one
       of two possible positions, or a display showing the state index 0 or 1.
\bibitem{canonical} The assumption of a canonical distribution is consistent with the
       argument of Popescu, Short, and Winter\cite{Popescu} that such a (mixed) state 
       derives from a typical pure state of the universe.
\bibitem{conventional} The decoherence arguments used here 
      follow from standard quantum theory with no added assumptions, and its results 
      are equivalent to those obtained from the conventional formalism, well illustrated in 
      a seminal papers by Zeh\cite{Zeh.70} and Zurek\cite{Zurek.81} and a review by 
      Schlosshauer.\cite{Schloss.04}
\end{thebibliography}
\end{document}